\documentclass{PoS}

\usepackage{amsmath}
\usepackage{amsfonts}
\usepackage{amssymb}
\usepackage{mathrsfs}
\usepackage{bm}

\newcommand{\mlo}{M_{\text{lo}}}
\newcommand{\mhi}{M_{\text{hi}}}
\newcommand{\chn}[3]{{{}^{#1}{#2}_{#3}}}
\newcommand{\cs}[2]{\chn{#1}{S}{#2}}
\newcommand{\cp}[2]{\chn{#1}{P}{#2}}
\newcommand{\cd}[2]{\chn{#1}{D}{#2}}
\newcommand{\cf}[2]{\chn{#1}{F}{#2}}
\newcommand{\csd}{{\cs{3}{1}-\cd{3}{1}}}
\newcommand{\cpf}{{\cp{3}{2}-\cf{3}{2}}}
\newcommand{\mnda}{\overline{\text{NDA}}}

\graphicspath{{../}}

\title{Renormalization and power counting of chiral nuclear forces}

\ShortTitle{Power counting of nuclear ChEFT}

\author{\speaker{Bingwei Long}
\thanks{In collaboration with Chieh-Jen Yang (University of Arizona)}
\\
        Jefferson Laboratory\\
        E-mail: \email{bingwei@jlab.org}}

\abstract{I will discuss the progress we have made on modifying Weinberg's prescription for chiral nuclear forces, using renormalization group invariance as the guideline. Some of the published results are presented.}

\FullConference{The 7th International Workshop on Chiral Dynamics,\\
		August 6 - 10, 2012\\
		Jefferson Lab, Newport News, Virginia, USA}

\begin{document}

\section{Introduction}

The stage of low-energy nuclear physics, where typical momenta are around or smaller than the pion mass $m_\pi$, suits very well chiral effective field theory (EFT)~\cite{Bedaque:2002mn, Epelbaum:2008ga, Machleidt:2011zz}. This is because chiral symmetry of quantum chromodynamics (QCD) and its spontaneous breaking provides a viable organization principle of calculations--- called power counting--- based on the expansion in ratios of small momenta to masses of heavy particles that are integrated out of chiral EFT, including non-Goldstone bosons and baryonic excited states. The consistency of power counting is especially relevant in calculations where reliable theoretical uncertainties are important.

For processes involving two or more nucleons, one-pion exchange (OPE) introduces, in addition to $\mhi \sim 4\pi f_\pi \simeq 1$ GeV, a low-energy scale in denominators: $\mlo \sim 4\pi f_\pi^2/m_N \simeq 0.1$ GeV. The coexistence of these two mass scales in denominators makes it unreliable to power count $NN$ contact interactions solely by way of dimensional analysis: How does one decide whether a derivative 4$N$ operator is suppressed by $\mhi$ or enhanced by $\mlo$? We seek from the requirement of renormalization group (RG) invariance to constrain the possible choices one can make {\it a priori} for power counting 4$N$ operators. More specifically, I will discuss our efforts to modify Weinberg's power counting (WPC), in pursuing a more consistent version of chiral nuclear EFT in the sense of fulfilling renormalization group invariance~\cite{Long:2011qx, Long:2011xw, Long:2012ve}. In this talk, I will focus on $S$ and $P$ waves where OPE needs to be resummed. For different points of view regarding renormalization in a nonperturbative EFT, see Refs.~\cite{Birse:2005um, Epelbaum:2009sd}.

\section{$NN$ chiral EFT has (at least) two scales}

All the considerations regarding short-range interactions do not change our view towards long-range physics that is reflected by non-analytic functions on external momenta stemming from loop integrals. Those non-analytic functions can still be reliably estimated through dimensional analysis by assuming internal lines going near mass shells. To see how $\mlo$ arises, let us consider the relative size of the box diagram compared to OPE:
\begin{align}
V_{1\pi} &\equiv \frac{g_A^2}{4 f_\pi^2} \bm{\tau}_1\bm{\cdot}\bm{\tau}_2 \frac{\vec{\sigma}_1\cdot\vec{q} \vec{\sigma}_2\cdot\vec{q}}{q^2 + m_\pi^2} \, , \\
V_{box}/V_{1\pi} &\sim \frac{m_N}{4\pi f_\pi} \frac{k}{\mathcal{A} f_\pi} \, ,
\end{align}
where $k$ is the center-of-mass momentum and $\mathcal{A}$ is a numerical factor that is of $\mathcal{O}(1)$ in lower partial waves. The low-energy scale $\mlo = \mathcal{A} f_\pi \sim 0.1$ GeV is related to the strength of OPE; it is not explicitly shown in chiral Lagrangian, but emerges through infrared enhancement inherent in low-energy, multiple-nucleon processes. 

The coefficients of $NN$ counterterms with $2n$ derivatives has mass dimension $2n$, multiplied by a common factor shared by most nonrelativistic theories:
\begin{equation}
C_{2n} \sim \frac{4\pi}{m_N} [mass]^{-2n} \, .
\end{equation}
With $\mlo$ and/or $\mhi$ able to make up the mass dimensions of $C_{2n}$, we need guidelines beyond dimensional analysis to decide whether $\mlo$, $\mhi$, or some combinations of both are contributing. RG invariance is such a principle we can use to constrain the choices in power counting the size of $C_{2n}$. More specifically, RG invariance considered here refers to the invariance of $NN$ scattering amplitudes with respect to a shift of the cutoff ($\Lambda$) that confines the three-momenta of nucleonic intermediate states in $NN$-reducible diagrams, since renormalization of pion loops in irreducible diagrams can be dealt with in a fashion very similar to standard chiral perturbation theory for single-nucleon processes. Before moving onto more detailed discussions regarding renormalization, we note that Weinberg's scheme~\cite{Weinberg:1990-1991} essentially chooses the minimal value for $C_{2n}$:
\begin{equation}
C_{2n} \sim \frac{4\pi}{m_N} \mhi^{-2n} \, .
\end{equation}
Besides being the most economical--- in terms of the number of counterterms used for a given order--- this choice cannot be immediately seen to satisfy RG invariance.

For chiral EFT in which the break-down scale is around $0.5$ GeV, RG invariance is interpreted differently than those theories that are designed to be a ``final'' theory, e.g., QCD. The EFT amplitudes of course cannot have an ``essential'' cutoff dependence, such as linear or logarithmic divergence, or oscillations that fluctuate beyond the desired uncertainty level like the amplitude of an unregularized singular attractive potential.

The RG invariance of EFT should be scrutinized further. Even if the cutoff dependence vanishes, the residual cutoff dependence for a finite value of $\Lambda$ must vanish fast enough so that the cutoff error, which is one of the sources for theoretical uncertainties, does not destroy the accuracy power counting has already prescribed. Of our interest is WPC, in which the $\mathcal{O}(Q)$ corrections have long been deemed to be zero; thus, the theoretical uncertainty for LO is considered by WPC to be $\mathcal{O}(Q^2)$. It follows that the LO cutoff error should vanish at least as fast as $Q^2/\Lambda^2$ for $\Lambda \sim \mhi$:
\begin{equation}
T^{(0)}(Q; \Lambda) - T^{(0)}(Q; \infty) \lesssim \left(\frac{Q}{\Lambda}\right)^{2}
\, .
\label{eqn_WPCLOres}
\end{equation}
While this is the case for the triplet channels~\cite{Beane:2000wh, PavonValderrama:2007nu}, we showed that it is not for $\cs{1}{0}$~\cite{Long:2012ve}, which forces us to modify WPC for $\cs{1}{0}$ at subleading orders even though its LO satisfies RG invariance, with the issues of $m_\pi^2$ dependence ignored for the time being~\cite{Beane:2001bc}.

\section{Subleading orders of $NN$ scattering amplitudes}

In a strictly perturbative theories, such as those of $\pi \pi$ scattering or $\pi N$ scattering, it is straightforward to identify those counterterms that are necessary to renormalize corresponding loop diagrams. However, it gets much more difficult to do so for nonperturbative problems like $NN$ scattering. I will divide my report into two cases: the triplet channels and $\cs{1}{0}$. The fundamental reason for that is that OPE behaves as $1/r^3$ for $r \to 0$ in the triplet channels while as $1/r$ in the singlet channels.

\subsection{Triplet channels}

It has been shown in Ref.~\cite{Nogga:2005hy} that the singular attractions in $\cp{3}{0}$ and $\cp{3}{2}$ demand more counterterms than WPC already at LO. Further studies by us~\cite{Long:2011qx, Long:2011xw}, following the approach proposed in Ref.~\cite{Long:2007vp}, aimed to find the counterterms that are needed to renormalize two-pion exchanges (TPEs) when they are treated as perturbations on top of the nonperturbative LO amplitudes. (References~\cite{Valderrama:2009ei, Valderrama:2011mv} used similar methods but came to conclusions with some differences.) The basic technique is to investigate the superficial degrees of divergence of $\langle \psi^{(0)} | V_{2\pi} | \psi^{(0)} \rangle$--- the matrix element of TPEs between the LO scattering wave functions~\cite{Long:2007vp, Valderrama:2009ei}.

The resulting arrangement of subleading counterterms in the triplet channels can be very nicely summarized by what we call $\mnda$: the subleading counterterms are enhanced by the same amount as the LO counterterm so that the whole tower of counterterms with the same quantum number is shifted uniformly. While this means that WPC remains intact in $\csd$ and $\cp{3}{1}$, it requires an enhancement of $\mathcal{O}(\mhi^2/\mlo^2)$ to all counterterms in $\cp{3}{0}$ and $\cpf$.

\subsection{$\cs{1}{0}$}

\begin{figure}
\centering
\includegraphics[scale=0.4]{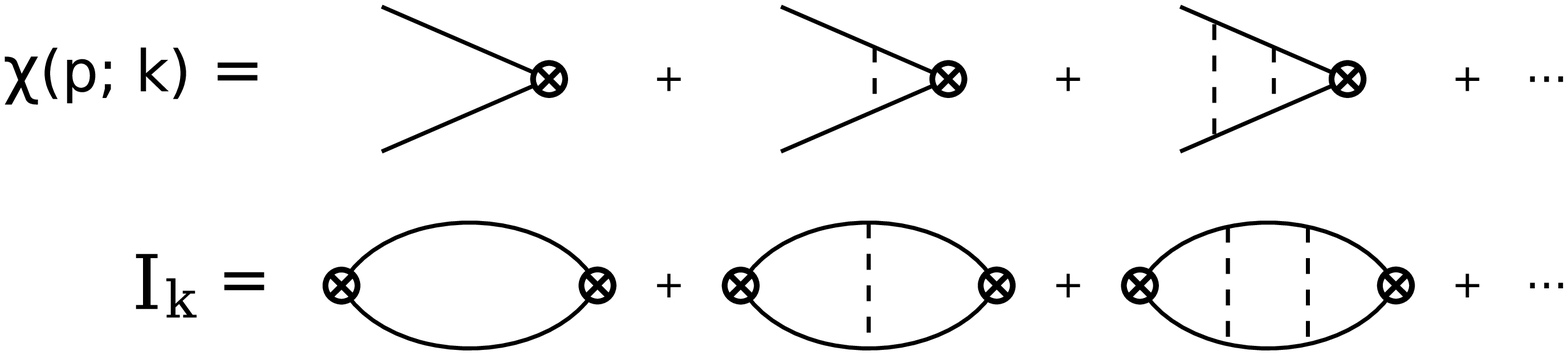}
\caption{Diagrammatic representation of $\chi(p; k)$ and $I_k$. Here the solid (dashed) lines represents the 
nucleon (pion) propagator, and the crossed circles do not represent any interaction.
\label{fig_I_k}}
\centering
\end{figure}

Using the technique developed in Ref.~\cite{Kaplan:1996xu}, we write the LO $\cs{1}{0}$ amplitude as
\begin{equation}
T^{(0)}(k) = T_Y(k) + \frac{{\chi(k)}^2}{C_R^{-1} - I_k^R + \mathcal{O}\left(\frac{m_N k^2}{4\pi \Lambda}\right) } \, ,
\label{eqn_TLOR}
\end{equation}
where $T_Y$ is the Yukawa-resummed amplitude, $I_k^R$ is the finite part of $I_k$, and $C_R$ is renormalized LO $\cs{1}{0}$ counterterm. The diagrammatic definitions of $\chi(k)$ and $I_k$ are shown in Fig.~\ref{fig_I_k}.

WPC prescribes that the first nontrivial corrections do not kick in until $\mathcal{O}(Q^2)$ (with LO labeled as $\mathcal{O}(Q^0)$, which means the (relative) theoretical uncertainty of $T^{(0)}$ is $\mathcal{O}(Q^2/\mhi^2)$. However, the residual cutoff dependence in Eq.~\eqref{eqn_TLOR} is $\mathcal{O}(Q^2/\mlo\Lambda)$, since $C_R^{-1} \propto \mlo$. In other words, WPC is being overly optimistic about the accuracy of the LO $\cs{1}{0}$ amplitude. This means a non-vanishing $\mathcal{O}(Q)$ can only be one insertion of the $C_2$ term--- the two-derivative $\cs{1}{0}$ contact operator  $C_2/2({p'}^2 + p^2)$.

Summarized in Table~\ref{table:PC} is our power counting for the two-nucleon sector in both singlet and triplet channels for $S$ and $P$ waves.

\begin{table}[bt]
\begin{tabular}{|c|c|}
  \hline
  $\mathcal{O}(1)$ & OPE,\; $C_\cs{1}{0}$,\;
  $\begin{pmatrix} C_\cs{3}{1} & 0 \\
  0 & 0
  \end{pmatrix} $,\;
  $C_\cp{3}{0}p' p$,\;
  $\begin{pmatrix} C_\cp{3}{2} p' p & 0 \\
  0 & 0
  \end{pmatrix}$ \\
  \hline
  $\mathcal{O}(Q)$ & $D_\cs{1}{0}({p'}^2 + p^2)$ \\
  \hline
  $\mathcal{O}(Q^2)$ & TPE0,\; $E_\cs{1}{0}{p'}^2 p^2$,\;
  $\begin{pmatrix} D_\cs{3}{1}({p'}^2 + p^2) & E_\text{SD}\,p^2 \\
  E_\text{SD}\,{p'}^2 & 0
  \end{pmatrix} $,\; \\
  & $D_\cp{3}{0}\,p' p({p'}^2 + p^2)$,\;
  $p' p \begin{pmatrix} D_\cp{3}{2}({p'}^2 + p^2) & E_\text{PF}\,p^2 \\
  E_\text{PF}\,{p'}^2 & 0
  \end{pmatrix} $, \\
  & $C_\cp{1}{1} p' p$,\; $C_\cp{3}{1}p' p$ \\
  \hline
  $\mathcal{O}(Q^3)$ & TPE1,\; $F_\cs{1}{0}{p'}^2p^2({p'}^2 + p^2)$ \\
  \hline
\end{tabular}
\caption{Power counting for pion exchanges, $S$ and $P$-wave counterterms up to $\mathcal{O}(Q^3)$. $p$ ($p'$) is the magnitude of the center-of-mass incoming (outgoing) momentum. The two-by-two matrices are for the coupled channels.\label{table:PC}}
\end{table}

\end{document}